\documentclass[11pt]{amsart}
\usepackage{amssymb,amsmath,latexsym,amscd,amsfonts}
\usepackage{graphics}
\usepackage{epsfig}
\usepackage{psfrag}
\def\ignore #1 {}

\def\hpic #1 #2 {\mbox{$\begin{array}[c]{l} \epsfig{file=#1,height=#2} \end{arr\
ay}$}}
\def\vpic #1 #2 {\mbox{$\begin{array}[c]{l} \epsfig{file=#1,width=#2} \end{arra\
y}$}}

\def\de{\partial}

\def\N{\mathcal N^{\text{vir}}}

\def\lrd{\!\stackrel{\leftrightarrow}{\partial}}
\def\dag{^\dagger}

\def\four{{\rm Four}}

\def\N{\mbox{{\bf N}}}

\def\R{\mathbb{R}}
\def\C{\mathbb{C}}

\def\Z{\mbox{{$\mathbb{Z}$}}}

\begin{document}

\title{Coisotropic Branes, Noncommutativity, and the Mirror Correspondence}
\author{Marco Aldi and Eric Zaslow}


\begin{abstract}
We study coisotropic A-branes in the sigma model on a four-torus by
explicitly constructing examples.  We find that morphisms between
coisotropic branes can be equated with a fundamental representation
of the noncommutatively deformed algebra of functions on the intersection.
The noncommutativity parameter is expressed in terms of the bundles
on the branes.  We conjecture these findings hold in general.
To check mirror symmetry, we verify that 
the dimensions of morphism spaces are equal
to the corresponding dimensions of morphisms between mirror objects.

\end{abstract}


\maketitle


\section{Introduction and Summary}

Kapustin and Orlov \cite{K} discovered coisotropic A-branes in
supersymmetric sigma models and speculated
about the role they played in the Fukaya category.\footnote{It has been
notoriously difficult to define the Fukaya category appropriately and to
describe its general objects -- see \cite{FOOO} \cite{Seidel}.}
In this paper, we study these
branes in examples,
by explicitly constructing the boundary superconformal field theories they
define.  Our examples reveal that the morphism spaces between coisotropic branes
are representations of a noncommutative algebra.  This algebra
is found to be the quantization
of the algebra of functions on the intersection of the coisotropic branes.
The representation is the fundamental representation of the algebra,
and generalizes the definition of Lagrangian intersections (where the
algebra is trivial at each intersection point). 
We conjecture this to be true in general.

More precisely, we study coisotropic branes on a four-torus $A$ appearing at
special points in K\"ahler moduli space.  Writing $E^\tau$ for $\R^2/\Z^2$ with
(complexified) symplectic form $\omega = \tau dx\wedge dy$ and $E_\tau$
for its mirror $\C/\Lambda_\tau,$  $\Lambda_\tau = \{n + m\tau\},$ we take
$A = E^\tau \times E^\tau$ with symplectic form $\pi_1^* \omega +
\pi_2^* \omega,$ where $\pi_i$ are the two projections.  The mirror
$\widetilde{A}$ is given by $\widetilde{A} = E_\tau \times E_\tau.$
When $\tau$ is an imaginary quadratic extension $E_\tau$ has
complex multiplication by $\alpha = n\tau,$ where $n$ is determined by
$\tau$.\footnote{For complex multiplication by $\widetilde{\alpha}$
we require $\widetilde{\alpha} \cdot \Lambda\subset\Lambda,$
so in particular $\widetilde{\alpha}\cdot 1 = m + n\tau,$
which is true if and only if we have complex multiplication by $\alpha = n\tau.$
The second lattice generator gives $(n\tau)\tau = -p + q\tau,$
whence $\tau = \frac{1}{2n}(q + i\sqrt{4pn - q^2}),$ where $4pn>q^2$
(otherwise $\Lambda \subset \R \subset \C$).} At such a point $\widetilde{A}$
attains a divisor $D=(z_2 = \alpha z_1)$ indicating a jump in the dimension of the
Picard group, i.e. a new line bundle $[D]$ appears ($[D]$ does not
extend to a bundle over a nontrivial complex family containing $\widetilde{A}$).
Correspondingly, $A$ should acquire a new A-brane, $(C,V),$ where
$C$ is coisotropic and $V$ is a $U(1)$ line bundle on $C.$

Now as $[D]$ is not (immediately) expressible in terms of pull-backs of line
bundles by the projections,
its mirror $C$ is thought to be a coisotropic brane, and in \cite{K} a Chern
class calculation was made to verify this point of view.  In that paper
the question of of defining Fukaya-type morphisms for coisotropic
branes was also raised. 

Here we explicitly construct the boundary
field theory, solve the boundary conditions, quantize and find the ground
states.  These ground states correspond to morphisms in the Fukaya-type
category of branes, and by studying coisotropic-coisotropic or
Lagrangian-coisotropic boundary conditions at the two endpoints
we can gain a concrete understanding of the mysterious
coisotropic branes in this example.  For example, ${\rm Hom}\left((C,-V^*),(C,V)\right)$
is found to
be the fundamental representation of the noncommutative algebra
of functions on $A_q,$ with
noncommutativity parameter $q$ determined by the curvature of $V.$
(Although the appearance of noncommutativity in the presence of curvature
is not entirely unexpected, the novelty here is that the boundary conditions
can be different on the two ends of the string, as opposed to a B-field.)
Finite-dimensionality of this representation is related to integrality
of the curvature form.

\section{Coisotropic Boundary Conditions}

In this section we will construct a supersymmetric sigma model with target
space a flat, symplectic four-torus on a Riemann
surface that is an infinite strip with two boundary components.  The
boundary conditions correspond to dual coisotropic A-branes
at the two ends.  We find mode expansions solving the equations of
motion and quantize to find the corresponding mode operators.
The explicit form of the Hamiltonian and supersymmetry generators $Q$
makes it easy to read off the states of $Q$-cohomology.

\subsection{Example:  $\tau = i$}

Consider the SUSY sigma-model with world-sheet
$\Sigma=[0,\pi]\times\R$ and target the flat square torus of real dimension four with
coordinates $X^\mu\sim X^\mu +2\pi$. The action is 
\begin{eqnarray*}
S = \frac{1}{4\pi} \int_\Sigma && \!\!\!\!\!\! dt ds   \left\{(\de_tX^\mu)^2  - (\de_s X^\mu)^2 + i
\theta^\mu_+ \de_- \theta_+^\mu + i \theta_-^\mu \de_+ \theta_-^\mu \right\} 
\cr\cr
&& + \frac{1}{2\pi}\int_{s=\pi} dt A_\mu\de_t X^\mu + \frac{1}{2\pi} \int_{s=0} dt A_\mu \de_t X^\mu
\end{eqnarray*}
where $\theta_\pm^\mu$ are real spinors and 
$$
\displaylines{
A_1=  \left\{ 
\begin{array}{ll} 
X^3     & 0 \le X^3 \le \pi                  \\ 
X^3 - 2\pi & \pi < X^3 < 2 \pi 
\end{array}
\right.
\,
A_2 = \left\{ 
\begin{array}{ll} 
- X^4 & 0\le X^4 \le\pi  \\ 
- X^4 + 2\pi & \pi< X^4 <2\pi 
\end{array}
\right.
\,
A_3= A_4=0}
$$
are abelian gauge fields. The boundary term is equivalent to two target-filling
D-branes with opposite $U(1)$ connection
$$
A = \pm \frac{i}{2\pi}[A_1dX^1+A_2 dX^2]
$$
and curvature
$$
F = dA = \pm \frac{1}{2\pi} (-dX^1 \wedge X^3 + d X^2 \wedge d X^4)
$$   
With respect to the symplectic form
$$
\omega=\frac{1}{2\pi}[d X^1 \wedge d X^2 + d X^3 \wedge dX^4]
$$
the A-brane condition of \cite{K} is satisfied
$$
F \omega^{-1} F  + \omega =0
$$
\subsubsection{Equations of motion and boundary conditions}
With respect to a variation 
$X^\mu \to X^\mu + \delta X^\mu $, the the boundary variation of the action is 
\begin{eqnarray*}
\delta S^B_{\rm bdry} &=& \frac{1}{2\pi} \int_{s=0} dt\, \{\delta X^1(-\de_t X^3 - \de_s X^1)+\delta X^2( \de_t X^4 - \de_s X^2)+ 
\cr
&&\qquad+\delta X^3( \de_t X^1 - \de_s X^3)+\delta X^4(-\de_t X^2 - \de_s X^4) \} + 
\cr
&+&  \frac{1}{2\pi}\int_{s=\pi} dt\,\{\delta X^1(-\de_t X^3+\de_s
  X^1)+\delta X^2(\de_t X^4 + \de_s X^2)+ 
\cr
&&\qquad+\delta X^3(\de_t X^1+\de_s X^3)+\delta X^4(-\de_t X^2+ \de_s X^4)\}
\end{eqnarray*}
Therefore, bosons must satisfy the wave
equation $(\de_s^2-\de_t^2)X^\mu = 0$ with boundary conditions
$$
\displaylines{
s = 0 \, : 
\qquad
\de_s X^1 =  \de_t X^3 \,;
\quad 
\de_s X^2 = - \de_t X^4 \,; 
\quad 
\de_s X^3 = - \de_t X^1 \,; 
\quad 
\de_s X^4 = \de_t X^2
\cr   
s = \pi \, :
\qquad
\de_s X^1 =  - \de_t X^3 \,;
\quad 
\de_s X^2 = \de_t X^4 \,; 
\quad 
\de_s X^3 = \de_t X^1 \,; 
\quad 
\de_s X^4 = - \de_t X^2
}
$$
A fermionic variation $ \theta^\mu \to \theta^\mu + \delta \theta^\mu $ 
induces a boundary variation 
\begin{eqnarray*}
\delta S^F_{\rm bdry} = \frac{i}{4\pi}\int_{s=0} dt \left\{ -\theta_+^\mu \delta \theta_+^\mu + \theta_-^\mu \delta \theta_-^\mu \right\} 
+\frac{i}{4\pi} \int_{s=\pi} dt \left\{ \theta_+^\mu \delta \theta_+^\mu - \theta_-^\mu \delta \theta_-^\mu \right\}  
\end{eqnarray*} 
The equation of motion for the fermions is the Dirac equation $\de_\pm
\theta^\mu_\mp=0$ and to impose the boundary conditions we require
\cite{B} invariance under $N=1$ SUSY with generators 
$$
\delta X^\mu = i \epsilon (\theta_+^\mu
 + \theta_-^\mu) \,;
\qquad 
\delta \theta_\pm^\mu = - \epsilon \de_\pm X^\mu
$$
and deduce
$$
\displaylines{
s = 0 \, : 
\qquad
\theta_+^1 =  \theta_-^3 \,;
\qquad 
\theta_+^2 = - \theta_-^4 \,;
\qquad
\theta_+^3 = - \theta_-^1 \,; 
\qquad 
\theta_+^4 =  \theta_-^2
\cr
s= \pi \, :
\qquad
\theta_+^1 = -\theta_-^3 \,;
\qquad 
\theta_+^2 = \theta_-^4 \,;
\qquad
\theta_+^3 = \theta_-^1 \,; 
\qquad 
\theta_+^4 = - \theta_-^2
}
$$
Moreover, bosonic and fermionic boundary conditions are invariant under N=2 supersymmetry with generators
$$
\delta X^\mu = \epsilon \theta_-^\mu - \overline{\epsilon} \theta_+^\mu \,;\quad \delta \theta_+^\mu = i \epsilon \de_+ X^\mu \,; \quad \delta \theta_-^\mu = -i \overline{\epsilon} \de_- X^\mu
$$
and supercharge 
\begin{eqnarray*}
Q =  \frac{1}{2\pi} \int_0^\pi\!\!\!\! &ds& \!\!\!\! (\theta_+^1 - i\theta_+^2)(\de_+ X^1 + i \de_+ X^2) + (\theta_-^1 + i\theta_-^2)(\de_- X^1 - i \de_- X^2)+
\cr
&&+(\theta_+^3 - i\theta_+^4)(\de_+ X^3 + i \de_+ X^4)+(\theta_-^3 + i\theta_-^4)(\de_+ X^3 - i \de_- X^4)   
\end{eqnarray*}
\subsubsection{Mode expansions}
Under the change of variables
\begin{eqnarray*}
Y^1(t,s) & = & \frac{1}{\sqrt{2}}(X^1(t,s) + i X^3(t,s)) \,;
\qquad
Y^2(t,s) = \frac{1}{\sqrt{2}}(X^2(t,s)+ i X^4(t,s))
\cr
\eta^1_\pm(t,s) & = & \frac{1}{\sqrt{2}}(\theta_\pm^1(t,s) + i \theta_\pm^3(t,s)) \,;
\qquad 
\eta^2_\pm(t,s) = \frac{1}{\sqrt{2}}(\theta_\pm^2(t,s) + i \theta_\pm^4(t,s))
\end{eqnarray*}
the boundary conditions decouple as
\begin{eqnarray*}
s = 0  &:& 
\qquad
\de_s Y^1 = i \de_s Y^1 \, ;
\quad 
\de_s Y^2 = - i \de_s Y^2 \, ;
\quad 
\eta^1_- = i \eta_+^1 \, ;
\quad
\eta_-^2 =  - i \eta_+^2
\cr
s = \pi &:&
\qquad
\de_s Y^1 =  - i \de_t Y^1 \; ;
\quad
\de_s Y^2 = i \de_t Y^2 \; ;
\quad 
\eta^1_- = - i \eta_+^1 \, ;
\quad
\eta_-^2 = i \eta_+^2
\end{eqnarray*}
The corresponding conjugate momenta
\begin{equation}
\label{momenta}
P^k := \frac{\de {\mathcal L}}{\de\de_t Y^k} = \frac{1}{2\pi} \de_t
  \overline Y^k + \frac{i}{4\pi}(Y^k - \overline Y^k + (\cdots))(\alpha_k \delta(s) + \beta_k\delta(\pi-s))
\end{equation}
where $\alpha_1 = \beta_1 = - \alpha_2 = - \beta_2 = - 1$ and $(\cdots)$ stands for the possible $2\pi$ factor in the piecewise definition of the gauge fields. We adapt to our case the mode expansion of \cite{C}
$$
Y^k(t,s) = y_k + i \left [\sum_{n=1}^\infty a_{k,n}\, \zeta_n^k
    (t,s)-\sum_{m=0}^\infty b_{k,m}\dag\, \zeta^k_{-m}(t,s) \right]
$$ 
where 
$$
\zeta_n^k(t,s) :=  |n - \epsilon_k|^{ - \frac{1}{2}}
  \cos \left[ ( n - \epsilon_k ) s + \tan^{-1} \alpha_k \right] e^{ - i (n -
    \epsilon_k)t}  
$$
and $\pi \epsilon_k = \tan^{-1}\alpha_k +\tan^{-1} \beta_k$.
Notice that boundary conditions are satisfied and the functions $\zeta_n^k$ 
form a complete orthogonal system with respect to the inner product 
$$
\int_0^\pi
\frac{ds}{\pi}\,\overline \zeta^k_n \, \left[ i \lrd_t +
  \alpha_k \delta(s) + \beta_k \delta(\pi-s) \right]\,\zeta^k_m = \delta_{mn}{\rm sign}\left(n-\epsilon_k\right).
$$
Moreover, each $\zeta_n^k$ is orthogonal to the zero mode in the sense that
$$
\int_0^\pi \frac{ds}{\pi}[i \de_t + \alpha_k \delta(s) + \beta_k \delta(\pi-s) ] \zeta^k_n = 0
$$
Using these relations, one can invert the mode expansions 
\begin{eqnarray}
\nonumber
a_{k,n} & = & \int_0^\pi
\frac{ds}{\pi}\overline{\zeta^k_n}[ \lrd_t - i \alpha_k \delta(s) - i \beta_k \delta(\pi-s)] Y^k  
\\
\nonumber
b_{k,n}\dag & = & \int_0^\pi
\frac{ds}{\pi} \overline {\zeta^k_{-n}}[\lrd_t -i \alpha_k \delta(s) -i \beta_k \delta(\pi-s)] Y^k 
\\
\label{zeromode}
y_k & = &\frac{1}{\alpha_k + \beta_k}\int_0^\pi [i \de_t + \alpha_k \delta(s) + \beta_k \delta(\pi-s)] Y^k 
\end{eqnarray}
For the mode expansion of the fermions, we follow \cite{B} (sec 39.1.2.5) 
$$
\eta^k_{\pm} = \sum_{n \in\Z} \eta_{k,n} e^{ \pm i [ (n-\epsilon_k) (s \pm
  t) + \tan^{-1} \alpha_k]} \, ;
$$ 
\subsubsection{Quantization}
We impose the equal-time canonical commutation relations
$$
[Y^k(t,s),P^j(t,s')] = [\overline Y^k(t,s), \overline P^j(t,s')] = \left\{ \eta^k_\pm(t,s), \eta^j_\pm(t,s') \right\} = i \delta_{kj} \delta(s - s')
$$
from which, using (\ref{momenta}) and (\ref{zeromode}), we deduce 
\begin{eqnarray*}
[y_k,y_k\dag] = \frac{\pi i}{(\alpha_k + \beta_k)^2} \int_0^\pi \!\!\!\!\! &\!\!\!& \!\!\!\!\int_0^\pi ds ds' ( \alpha_k \delta(s') + \beta_k \delta(\pi-s'))  [ \overline P^k(t,s), \overline Y^k(t,s') ] +
\cr
 \!\!\!\!\!& \!\!\!&\!\!\! +( \alpha_k \delta(s) + \beta_k \delta(\pi-s)) [ Y^k(t,s), - P^k(t,s') ] = \frac {2 \pi}{\alpha_k +\beta_k}
\end{eqnarray*}
Similarly,
$$
[b_{k,n},b_{k,m}\dag] = \delta_{nm} = [a_{k,n},a_{k,m}\dag]
$$
The nontrivial fermionic anticommutators are (see e.g \cite{B} sec 39.1.2.25)
$$
\{\eta_{k,n},\eta\dag_{k,m}\} = \delta_{nm} 
$$ 
The mode expansion for the Hamiltonian is 
\begin{eqnarray*}
H &=& \frac{1}{\pi}\int_0^\pi ds\, |\de_t Y^k|^2+|\de_s
Y^k|^2 + \frac{i}{2}(\overline \eta_+^k\lrd_s\eta^k_+
-\overline \eta_-^k \lrd_s\eta_-^k)=
\cr
&=&\sum_{n=1}^\infty (n+\frac{1}{2}) a\dag_{1,n} a_{1,n}  
+\sum_{m=0}^\infty (m-\frac{1}{2}) b_{1,m}\dag b_{1,m} +\sum_{n=0}^\infty
(n-\frac{1}{2})a\dag_{2,n} a_{2,n}+
\cr
&+& \sum_{m=1}^\infty (m+\frac{1}{2})b_{2,m}\dag
b_{2,m} + \sum_{n=1}^\infty (n+\frac{1}{2})\eta\dag_{1,n}\eta_{1,n}+\sum_{m=0}^\infty
(m-\frac{1}{2})\eta_{_{1,-m}}\eta\dag_{_{1,-m}}+
\cr
&+&\sum_{n=0}^\infty
(n-\frac{1}{2}) \eta\dag_{2,n}\eta_{2,n}+\sum_{m=1}^\infty
(m+\frac{1}{2})\eta_{_{2,-m}}\eta\dag_{_{2,-m}}
\end{eqnarray*}
Notice that the  zero-modes commute with the
Hamiltonian. Using $\sqrt 2 y_1 = x_1 + i x_3$,  the commutation 
relation for the bosonic zero-modes can be rewritten as
\begin{equation}
\label{xcomms}
[x_1,x_3] = - i \pi
\end{equation}
Therefore, we can think of  
$$
p_1:= - \frac{x_3}{\pi}
$$ 
as the conjugate momentum for $x_1$ and use it to label ground
states. As in \cite{C}, compactness of the target implies the
existence of finitely many ground states. In fact, on the one hand,
the zero-mode wave function 
$$
\langle x_1|p_1\rangle=e^{i p_1 x_1}
$$
has to remain single-valued under $x_1\rightarrow x_1+2\pi$ and this
implies $p_1\in\Z$. On the other hand, $x_3\rightarrow
x_3+2\pi$ implies  $p_1\rightarrow p_1-2$ so that there are only two
independent zero-mode wave functions. A similar analysis can be
repeated for the other zero-mode $y_2$ so that overall there are four
ground states of the form $|p_1,p_2\rangle$ corresponding to the
values $p_1=0,1$ and $p_2= 0,1.$
Using the relation $2H=\{Q,Q\dag\}$, we see that excited states are
killed by the BRST 
operator.  In particular fermionic modes do not contribute to the
cohomology as, due to the fractional mode expansion, they have either positive
or negative energy.  Therefore, the dimension of the BRST cohomology
ring is 4.

Another point of view on the space of ground states is as follows.  Define
$U = e^{ix_1}$ and $V = e^{i\pi p_1} = e^{i x_3}.$  Then the relation
(\ref{xcomms}) together with periodicity defines the algebra
$$
U^2 = V^2 = 1,\quad UV = - V U.
$$
This algebra is the noncommutative two-torus!
The fundamental representation is two dimensional
and can be constructed from the $x_1$-momentum eigenvectors
$e_1 = \vert p_1 = 0\rangle, e_2 = \vert p_1 = 1\rangle,$
with $U$ and $V$ acting as matrices
$$
U=\left(\begin{array}{cc}0&1\\1&0\end{array}\right),\quad V = 
\left(\begin{array}{cc}1&0\\0&-1\end{array}\right).
$$
Note that $U$ increases the $p_1$ eigenvalue by $1$ (mod 2) and we
have made $V$ diagonal by the basis choice.
The states $x_2$ and $x_4$ define another copy of
this algebra.  The ground states
$|i,j\rangle$ are identified with $e_i\otimes e_j.$
The dimension of the representation of the ground state
algebra, $4,$ agrees with the mirror calculation performed in section
\ref{secbmod}.

\subsection{Coisotropic-Coisotropic and Lagrangian-Coisotropic Examples}

We consider some examples along the general lines of the previous
section, abbreviating the the analysis somewhat.

\subsubsection{Coisotropic-Coisotropic Boundary Conditions}

If we fix the symplectic structure to be the standard one $\omega$, then a skew-symmetric $4 \times 4$ matrix $F=(a_{ij})$ with integer entries is the curvature of a coisotropic A-brane if and only if
$$
F \wedge \omega=0 \,;
\qquad
{\rm Pfaff}(F)=\frac{1}{2} \int F \wedge F = 1
$$ 
Notice that if $F$ is of this form then so is $-F,$
and one can study the open string states from $-F$ to $F$ as before. Therefore, our computation works for any such pair $(-F,F)$, the results depending only on the determinant of the curvature. When the curvatures $F$, $G$ of the branes at the endpoints are not multiples of each other, we need to modify the mode expansion. Consider for example the case 
\begin{eqnarray*}
&&s = 0:\quad  F = -J = dX^1\wedge dX^3 - dX^2 \wedge dX^4  
\\
&&s = \pi:\quad G = K = - dX^1\wedge dX^4 - dX^2 \wedge dX^3
\end{eqnarray*}
This notation is suggested by the familiar quaternionic relations
$$
I^2=J^2=K^2=-1 \, ;
\qquad 
IJ=K
$$
where $I=\omega^{-1}g$ (here $g$ is the identity matrix).
It is convenient to introduce the quaternionic field
$$
Q(s,t) = X^1(s,t) + I X^2(s,t) + J X^3(s,t) + K X^4(s,t)
$$
with equation of motion  $(\de_s^2- \de_t^2)Q=0$ and boundary conditions
$$
\de_s Q(0,t) = - J \de_t Q(0,t) \, ;
\qquad
\de_s Q(\pi,t) = K \de_t Q(\pi,t)
$$ 
We can solve the boundary conditions with the mode expansion
$$
Q(s,t) = q_0 + \sum_{n\in \Z} e^{-I(n+\frac{1}{4})s} e^{K(n+\frac{1}{4})t} q_n
$$
In analogy with the previous example,
$$
q_0 = x_1 + I x_2 + J x_3 + K x_4  = \frac{J + K}{2}\int_0^\pi ds [\de_t - J \delta(s) - K \delta(\pi-s) ] Q(s,t)
$$
Using
\begin{eqnarray*}
P^1 &=& \frac{1}{2\pi} \de_t X^1 + \frac{1}{2\pi} X^3 \delta (s) + \frac{1}{2\pi} X^4 \delta(\pi-s) \, ;  
\quad
P^3 = \frac{1}{2\pi} \de_t X^3 
\cr
\cr
P^2 &=& \frac{1}{2\pi} \de_t X^2 - \frac{1}{2\pi} X^4 \delta (s) + \frac{1}{2\pi} X^3 \delta(\pi-s) \, ; 
\quad
P^4 = \frac{1}{2\pi} \de_t X^4 
\end{eqnarray*}
we deduce
\begin{eqnarray*}
x_1 &=&  \frac{1}{2}\int_0^\pi \left(-2 \pi P^3 - 2 \pi P^4 + (X^1 - X^2 ) \delta(s) + (X^1  + X^2)\delta (\pi-s) \right)
\cr 
x_2 &=&  \frac{1}{2}\int_0^\pi \left( 2 \pi P^3 - 2 \pi P^4 + ( X^1 + X^2) \delta(s) + (-X^1 + X^2) \delta(\pi-s) \right)
\cr
x_3 &=& \frac{1}{2}\int_0^\pi \left( 2 \pi P^1 + 2 \pi P^2  \right) 
\cr
x_4 &=& \frac{1}{2}\int_0^\pi \left( 2 \pi P^1 - 2 \pi P^2  \right) 
\end{eqnarray*}
The canonical commutation relations imply the non vanishing-commutators 
\begin{eqnarray*}
[x_1, x_3] &=&  \pi i \,; \quad [x_1, x_4] = \pi i
\cr
[x_2, x_3] &=&  \pi i \,; \quad [x_2, x_4] = - \pi i
\end{eqnarray*}
Note that $[x_j , x_k] = ((G-F)^{-1})_{jk}$. As we will see in
sec. \ref{general}, these relations imply that there are two ground
states.

\subsubsection{Lagrangian-Coisotropic Boundary Conditions}
\label{lagcois}

The case where only one of the branes is strictly coisotropic can be treated in a similar way. As a simple example, suppose that the bosonic part of the action is
$$
S=\frac{1}{4\pi}\int_\Sigma dt ds \{ \de_t X^\mu)^2 - ( \de_s X^\mu)^2 \} + \frac{1}{2\pi} \int_{s=\pi} A_{\mu} \de_t X^\mu dt
$$
and we impose boundary conditions
\begin{eqnarray*}
s = 0   &:& X^2 = 0 \, ; \quad \de_s X^1 = 0 \, ; \quad X^4 = 0 \, ; \quad \de_s X^3 = 0
\cr
s = \pi &:& \de_s X^1 = - \de_t X^3 \,; \quad \de_s X^2 = \de_t X^3\, ; \quad \de_s X^3 = \de_t X^1 \, ; \quad \de_s X^4 = - \de_t X^2 
\end{eqnarray*}
Since $X^2$ and $X^4$ have no interesting zero modes, we focus on complex solutions  
$$
Y(s,t) = \frac{1}{\sqrt 2} (X^1(s,t) + i X^3(s,t))
\quad{\rm s.t.}\quad
\de_s Y(0,t)= 0 \,; \quad \de_t Y(\pi,t)= i \de_s Y(\pi,t) 
$$
We use the expansion
$$
Y(s,t) = y + \sum_{n\in \Z} a_n \cos \left[ \left( n - \frac{1}{4} \right) s \right] e^{-i (n - \frac{1}{4})}
$$
and in analogy with the above computation
$$
[y, y\dag] = [\int_0^\pi ds \{i\de_t + \delta(s-\pi)\} Y(s,t) , \int_0^\pi ds\{-i \de_t + \delta(s-\pi)\} \overline Y(s,t)] = -2 \pi
$$
Therefore, $[x_1,x_3] = - 2 \pi i$ and we have a single bosonic state.

\subsection{Noncommutativity in the General Case}
\label{general}

For the general torus $\R^{2d}/(2\pi\Z)^{2d},$ we
consider coisotropic branes with bundles ${\mathcal E}_1$ at $s=0$
and ${\mathcal E}_2$ at $s=\pi,$ with curvatures
defined by the integer, skew-symmetric
matrices $F_{ij}$ and $G_{ij},$ respectively. We will now argue that
the zero modes obey the following commutation relations: 
\begin{equation}
\label{nctorus}
[x^k,x^l] = 2\pi i A^{kl},
\end{equation}
where $A = G-F$ is also an integer-valued, skew-symmetric $2d\times 2d$ matrix,
which we assume to be invertible,\footnote{This assumption is violated, for example,
when one considers open strings between a coisotropic brane and itself.  In
such an example, there are fermionic zero modes and the ground state
sector is not simply given by the zero-mode algebra.  A unified treatment
of coisotropic branes should yield a mathematical cohomology theory matching
BRST cohomology, as Gualtieri's $d_L$ operator does
for the endomorphisms of coisotropic branes {\sl qua} generalized
complex submanifolds \cite{Gual}.} while
$A^{kl}$ are the components of the inverse matrix:  $A^{kl}A_{lm}=\delta^k{}_m.$
To see this, suppose that $\zeta$ is a solution to the boundary
problem
$$
(\de_s^2 - \de_t^2) \zeta(s,t) = 0 \,;\quad \de_s \zeta(0,t) = F \de_t
\zeta(0,t) 
\,;\quad \de_s \zeta(\pi,t) = G \de_t \zeta(\pi,t)  
$$
and moreover that its $t$-antiderivative $\psi$ is also a solution to
the same problem (this is reasonable since $F$ and $G$ are constant matrices). Under these assumptions,
\begin{equation}
\label{innerprod}
\int_0^\pi ds [ \de_t + F \delta(s) - G \delta(\pi-s)] \zeta = \de_s
(\psi(\pi)-\psi(s))+F \de_t \psi (0) - G\de_t\psi(\pi)=0  
\end{equation}
i.e. we have an inner-product which makes all possible oscillators 
orthogonal to the zero-modes. Therefore, the components of the zero-mode of the field $X$ are
$$
x^l = \frac{a^{ml}}{2} \int_0^\pi 4\pi P^m + (f_{mn}
\delta (s) -g_{mn} \delta (\pi-s)) X^n  
$$
We conclude 
\begin{equation}
\label{commutator}
[x^k,x^l] = a^{jk} a^{ml} [2\pi P^j + \frac{1}{2}
  a_{mj} X^m, 2\pi P^m + \frac{1}{2} a_{jm} X^j] =  2\pi i a^{kl}       
\end{equation}
(Similarly, one can compute the zero-mode algebra when the brane with
charge $F$ is replaced with the Lagrangian brane of \ref{lagcois}. It
is enough to discard even-numbered coordinates and set $F=0$ in
(\ref{innerprod}) and (\ref{commutator}). Moreover, a rotation of the
coordinates sends a coisotropic brane to another coisotropic one so
that the same computation works for a
linear Lagrangian placed at $X^{2k}(0)= m_k X^{2k-1}(0)$ where $m_k\in
\mathbb Q$, $k=1,\ldots,d$). We assume that our coordinates are $A$-symplectic in the sense that
$A_{ab} = 0,$ $a,b = 1...d$ and similarly for $A_{d+a,d+b}.$  In fact, define
$u^a = x^a,$ $v^a = x^{d+a}.$    
Therefore, $A$ is defined by the
$d\times d$ integer matrix $N,$ i.e. $A$ has block-anti-diagonal form
$$A = \left( \begin{array}{cc}0&N\\-N^T&0\end{array}\right).$$
Note
$$A^{-1} = \left(\begin{array}{cc}0&-N^{-T}\\N^{-1}&0\end{array}\right).$$
Then from the relations (\ref{nctorus}), we find the corresponding momenta
and relations
$$p_j = \frac{1}{2\pi} x^k A_{kj}\qquad [p_j,p_k] = -\frac{i}{2\pi}A_{jk}$$
(or $p_{u^a} = -\frac{1}{2\pi}v^b N_{ab}$ and $p_{v^a} = \frac{1}{2\pi}u^b N_{ba}$).
Note that the Dirac quantization condition applied to the two-dimensional
sub-torus defined by the coordinate pair $(jk)$ says $(2\pi)^2 (-1/2\pi) A_{jk}
= 2\pi n,$ where $n\in\Z,$ i.e. $A$ is an integer-valued matrix.  So our relations
are consistent.
Since we can rewrite the momenta as coordinates, the lattice
translation in the $k$-direction, $\vec{x} \rightarrow \vec{x} + 2\pi e_k$
implies the corresponding periodicity in the momentum lattice
$$\vec{p} = p_i e^i  \sim (p_i + A_{ki})e^i, \qquad k = 1,...,2d,$$
or
$$\vec{p}_u = p_{u^a}e^a
\sim (p_{u^a}+N_{ab})e^b,\qquad
\vec{p}_v = p_{v^a}e^{d+a} \sim (p_{v^a}+N_{ba})e^{d+b}.$$

Now define the operators
$$U_a = e^{iu^a},\qquad V_a = e^{iv^a},$$
well-defined by the $2\pi$-periodicity of the coordinates.
By the form of $A,$ the $U$ operators commute with each other,
as do the $V$ operators (so we can diagonalize $u$-momenta or
$v$-momenta, but not both).
Now we interpret the operator
$U_a$ as adding a unit of $p_{u^a}$ momentum,
whereas $V_a$ adds a unit of $p_{v^a}$ momentum.
This imposes the following relations among the $U$ and $V$
\begin{eqnarray*}
U_{1}^{N_{1a}}U_{2}^{N_{2a}}...U_{d}^{N_{da}} &=& 1\\
V_{1}^{N_{a1}}V_{2}^{N_{a2}}...V_{d}^{N_{ad}} &=& 1\qquad a = 1,..,d
\end{eqnarray*}

To write this in a more convenient form, define
$U^\alpha = U_1^{\alpha_1}U_2^{\alpha_2}...U_d^{\alpha_d}$
for $\alpha \in \Z^d.$  Define $N_{a,\bullet} = (N_{a1},N_{a2},...N_{ad})$
to be the $a$-th row of $N;$ similarly define $N_{\bullet,a}$ to be the $a$-th
column.  Then the relations of the zero-mode algebra can now be written
\begin{eqnarray}
U^{N_{a,\bullet}} &=& 1,\\
V^{N_{\bullet,a}} &=& 1,\\
U_aV_b &=& e^{2\pi i N^{ba}} V_bU_a,\\
U_a U_b &=& U_b U_a,\\
V_a V_b &=& V_b V_a, \qquad a,b = 1,...,d. 
\end{eqnarray}
This is a noncommutative torus defined by the matrix $N.$
Note that since $N$ is integer, the components of $N^{-1}$ are
rational, so all the noncommutativity parameters are roots of unity.
The zero modes naturally define conjugate finite-dimensional
$u$-momentum and $v$-momentum eigenstate representations, $H_u$ and $H_v,$
by construction.  The dimension of the representation corresponds
to the index of the sublattice in $\Z^d$ defined by the vectors $N_{a,\bullet}$
(or for the conjugate representation $N_{\bullet,a}$), so
$${\rm dim}H_u = {\rm dim}H_v = {\rm det} N = {\rm Pfaff}(A) = \frac{1}{d!}
\int c_1(\mathcal{E}_1^*\otimes\mathcal{E}_2)^d.$$

We can now speculate about the general case.  Noncommutativity arose
from the nonexistence of solutions to the equations of motion that were linear
in time.  As a result, there were no separate conjugate momenta
to the zero-mode oscillators -- they formed their own momentum conjugates
and did not simultaneously commute.
This argument also depended on the decoupling of these
modes from the oscillators of the open string. 
Let us assume such a decoupling
in general\footnote{This assumption would follow from an
orthogonal basis of solutions to the appropriate Laplace equation (equations
of motion) with boundary conditions determined by the curvatures,
with respect to an appropriately defined inner product.} and study two A-branes $(C_1,F_1)$ and $(C_2,F_2)$
at the left and right boundary conditions.  Here $C_i$ can be coisotropic or
Lagrangian and $F_i$ is the curvature of a connection $A_i$ on a bundle
${\mathcal E}_i\rightarrow C_i.$
We study the zero modes $x^i.$ 
The boundary conditions require $x^i$ to lie at the intersection
$C_1\cap C_2,$ and we would like to compute the zero-mode
algebra. In the case of transverse Lagrangian-Lagrangian
intersection, the zero modes carry no degrees of freedom so their algebra
is trivial, except that there is a copy of it at each intersection point.  We think of
this case thus trivially falling into the noncommutative picture.  Now consider
the case where at least one brane is coisotropic.
We assume that the curvature
$F = F_2 - F_1$ on ${\mathcal E}_1^*\otimes{\mathcal E}_2$ is
nondegenerate, so there are no fermionic zero-modes and the BRST cohomology
is equal to the representation of the zero modes.  The natural generalization
of (\ref{nctorus}) is that $F^{-1}$ is the Poisson structure of a
(Moyal) noncommutativization
of the algebra of functions on the intersection $C_1\cap C_2.$  
In the examples we computed, this algebra had a
``fundamental'' representation $H,$ finite-dimensional due
to the integrality of the curvature two-form,
with $${\rm dim}_\C H = \frac{1}{d!}\int_{C_1\cap C_2} c_1^{d}({\mathcal E}),$$
where $d$ is half of the real dimension of the intersection.
We do not know whether finite dimensionality or the
dimension formula holds in the general case.
We conjecture this to be the case.

\section{B-Model Calculations}
\label{secbmod}

We now wish to describe the mirror B-model calculations of what
we have done in previous sections.  The cohomology of a holomorphic
bundle over a complex torus can be constructed explicitly.  We follow the
beautiful treatment given in \cite{Pol}.

\subsection{Mirror geometry}
Let $\widetilde{A} = E_\tau \times E_\tau$ be the product of two
elliptic curves with modular parameter $\tau = \tau_1 + i \tau_2$ and
let $z$, $w$ be complex coordinates on the first and second factor
respectively. 
With respect to a standard symplectic basis $x_1, y_1, x_2, y_2$ for
$H^1(\widetilde{A},\Z)$, the complex structure is block diagonal and on each $2\times 2$ factor
is given by 
$$
\mathcal J
=
-\frac{1}{\tau_2}
\left(
\begin{matrix}
\tau_1 & |\tau|^2 \\
-1     & -\tau_1
\end{matrix}
\right)
=
\left(
\begin{matrix}
1 & \tau_1 \\
0 & \tau_2
\end{matrix}
\right)
\left(
\begin{matrix}
0 & -1 \\
1 & 0
\end{matrix}
\right)
\left(
\begin{matrix}
1 & \tau_1 \\
0 & \tau_2
\end{matrix}
\right)^{-1}
$$
The Neron-Severi group $NS(\widetilde{A})$ has three generators $L_i$ such that 
\begin{eqnarray*}
c_1(L_1) = \frac{1}{\tau_2}{\rm Im}(({\bar \tau} x_1 - y_1)(\tau x_1 - y_1)) & = & x_1 y_1 
\cr
c_1(L_2) = \frac{1}{\tau_2}{\rm Im}(({\bar \tau} x_2 - y_2)(\tau x_2 - y_2)) & = & x_2 y_2 
\cr
c_1(L_3) = \frac{1}{\tau_2}{\rm Im}(({\bar \tau} x_1 - y_1)(\tau x_2 - y_2)) & = & x_1 y_2 - x_2 y_1 
\end{eqnarray*}
for generic $\tau$. The first two correspond to pull back divisors
under the two projections, while the third appears due to the
existence of the diagonal divisor. When $E_\tau$ admits complex
multiplication by $\alpha$ there is an extra bundle $L_4,$ due to the
divisor $ w = \alpha z,$ with first Chern class
multiple of  
$$ 
{\rm Re} (({\bar \tau} x_1 - y_1)(\tau x_2 - y_2)) = (\tau_1^2 + \tau_2^2) x_1 x_2 + y_1 y_2 - \tau_1 (x_1 y_2 + y_1 x_2) 
$$
It is straightforward to compute the Chern character of the mirror object via Fourier transform. For example, Kapustin and Orlov \cite{K} observed that for $\tau=i$ 
$$
ch({\rm Four}(L_4)): = \int ch(L_4) ch(P) = 1- l_1l_2 + y_1 y_2 + l_1 y_1 l_2 y_2 
$$
where $l_1$, $l_2$ are coordinates dual to $x_1$, $x_2$, integration is performed with respect to the form $x_1 x_2$ and $P$ is the Poincar\'e line bundle such that $c_1(P)= x_1 l_1 + x_2 l_2$. In fact, in coordinates 
$$
l_1 = \frac{1}{2\pi} dX^1\, , \quad l_2 = \frac{1}{2\pi} dX^3\, ,\quad
y_1 = \frac{1}{2\pi} dX^2\, , \quad y_2 = \frac{1}{2\pi} dX^3\, ;
$$
this coincides with the Chern character of the coisotropic brane $J$
used in our first example. In a similar way, we interpret the
coisotropic brane $-J$ as the Fourier transform of $-[L^*_4]$ i.e. the
opposite in $K_0(\widetilde{A})$ of the line bundle dual to $L_4$. With this
identification, one can apply the standard B-model techniques (see
e.g. \cite{W}, \cite{B}) and use the Gothendieck-Riemann-Roch theorem to
compute  
$$
\dim {\rm Ext}^0(-[L_4^*], L_4) = - \chi (L_4^*, L_4) = \int_{\widetilde{A}}
ch^2(L_4) = 4 \, ,
$$
in agreement with the A-model calculation. In addition, these elements
can be identified with $H^0(L_4^2),$ which naturally carries a representation
of the Heisenberg group \cite{Pol}, i.e. the very same noncommutative algebra!

More generally, consider two coisotropic A-branes $\mathcal E_1,
\mathcal E_2$ as in section  \ref{general}. The mirror brane of $\mathcal E_1$ has Chern
character  
\begin{eqnarray*}
ch(\four(\mathcal E_1)) & = & \int ch(\mathcal E_1) ch(P) = 
\cr
&=& -f_{13} + f_{23} x_1 y_1 + x_1 x_2 - f_{34} x_1 y_2 - f_{12} y_1
x_2 + y_1 y_2 + f_{14} x_2 y_2 +
\cr
&\,&+ f_{24} x_1 y_1 x_2 y_2   
\end{eqnarray*}
and similarly for $\mathcal E_2$. It follows that 
\begin{eqnarray*}
\chi(\four(\mathcal E_1),\four(\mathcal E_2)) & = & \int ch((\four(\mathcal
E_1))^*) ch({\four(\mathcal E_2)}) = 
\cr
& = & {\rm Pfaff}(F)+ {\rm Pfaff}(G) + \epsilon^{ijkl} f_{ij} g_{kl} = {\rm Pfaff}(G-F)\end{eqnarray*}
where $\epsilon^{ijkl}$ is the Levi-Civita symbol. We conclude that
the number of BRST states computed in section \ref{general} is
compatible with mirror symmetry. 

\section{Conclusion}

Several issues remain to be understood.  First of all, one should be able to
adapt all of these computations to the case of $\tau \neq
i$. Secondly, the relation between Heisenberg group representation 
and the noncommutativity of the A-model should be clarified. 
Finally, it would be interesting to interpret these computations
from the point of view of the analogue of the K\"unneth theorem for
triangulated categories proved in \cite{BLL}. Kontsevich has suggested that
coisotropic branes should be considered as elements
in the idempotent completion of the tensor square of
the Fukaya category of $E^\tau$.  From this point of view, one only
needs to understand the idempotents defining coisotropic summands,
since the morphisms (and compositions!) of summands can be
computed from these.

In summary, it has been anticipated that coisotropic branes appear as
summands in 
the category generated by Lagrangian branes and their shifts and sums, but
a more direct treatment of these objects has been lacking.  We hope that
the description of morphisms in terms of representations of noncommutative
algebras will help clarify the role of coisotropic branes in the
Fukaya category 
and perhaps be relevant to the study of their deformations.\footnote{The
deformation theory of classical coisotropic submanifolds has
already proven quite interesting \cite{O,OP}.}   Eventually, one would like
to know what an A-brane is!

\vskip 0.2in
\centerline{\bf Acknowledgments}
\vskip 0.1in
We would like to thank Kentaro Hori, Anton Kapustin, Alexander Polishchuk and
Boris Tsygan for helpful discussions.
The work of EZ has been supported in part by an
Alfred P. Sloan Foundation fellowship, by NSF
grants DMS--0072504 and DMS--0405859, and by the
Clay Senior Scholars program.  EZ also thanks
the School of Mathematics, Statistics, and Computer
Science at Victoria University, Wellington, while both
authors thank The Fields Institute, where some
of this work took place.


\vskip 0.2in
{\scriptsize
{\bf Marco Aldi,} Department of Mathematics, Northwestern University,
2033 Sheridan Road, Evanston, IL  60208 (m.aldi@math.northwestern.edu)\\
{\bf Eric Zaslow,} Department of Mathematics, Northwestern University,
2033 Sheridan Road, Evanston, IL  60208 (zaslow@math.northwestern.edu)
}


\end{document}